\documentclass{article}[11pt]

\usepackage{amsfonts,amsmath,amssymb,graphicx,epsfig,color,setspace,hyperref}

\def\be {\begin{equation}}
\def\ee {\end{equation}}
\def\bea {\begin{eqnarray}}
\def\eea {\end{eqnarray}}
\def\bc {\begin{center}}
\def\ec {\end{center}}
\def\bfg {\begin{figure}}
\def\efg {\end{figure}}
\def\bi {\begin{itemize}}
\def\ei {\end{itemize}}

\def\beq{\begin{equation}}
\def\eeq{\end{equation}}
\def\br{\begin{eqnarray}}
\def\er{\end{eqnarray}}
\newcommand{\eel}[1] {\label{#1}\end{equation}}

\newcommand{\bdm}{\begin{displaymath}}
\newcommand{\edm}{\end{displaymath}}

\begin{document}

\onehalfspacing

\thispagestyle{empty}

\begin{center}

{\Large \bf  Timeless state of gravity:\\ Black hole universal clock}

\bigskip
\bigskip
\bigskip

{ \large   Ahmed Farag Ali$^{\triangle \dagger}$\footnote{\tt Email: ahmed.farag.ali@gravitonworld.org \\ ahmed.ali@fsc.bu.edu.eg}}

\bigskip
\bigskip

{\em $^{\dagger}$GRAVITON, San Mateo, CA 94401, USA} \\
{\em $^\triangle$Dept. of Physics, Benha University, Benha 13518, Egypt} \\

\bigskip
\bigskip
\bigskip

{\bf Abstract} \\
\noindent
\end{center}
We investigate  Rindler's frame measurements. From its perspective, we found a geometric/gravitational interpretation of speed of light, mass and uncertainty principle. This can be interpreted as measurements of a black hole universal clock. This lead to an emergence of a timeless state of gravity in a mathematically consistent way. In other words, space my be a frozen time.
\bigskip
\bigskip
\bigskip

\begin{center}
\noindent
\end{center}

\newpage

\setcounter{page}{1}

\section{Clock Postulate and Global Reference Frame}

One of the widely known incompatibilities between special relativity and general relativity is how each theory sees red-shift. In special relativity, the moving observers notice clocks rate change by a Lorentz factor $\gamma$. But stationary observers share the same clocks rate even if they are at different spatial positions as the electromagnetic wave frequency is only a source-dependent property. Therefore, if those stationary observers notice any shift in the wavelength, then it is justified only if the observer's clock ticks with different rates at different spacetime points, which is not compatible of special relativity. In fact the general theory of relativity justifies the change in clock rate as a response to the change in the gravitational potential from point to another, i.e., gravitational red-shift is a property of general covariance, which is global principle of general relativity. That is why A. Einstein realized that special theory of relativity and the equivalence principle hold locally, not globally \cite{Einstein:1911}. This is clearly manifested in accelerated Rindler spacetime with metric $ds^2=-r^2dt^2+dr^2$ \cite{Rindler:1966zz} such that the observer's clock rate is determined by $R(r)=\big(1-2V(r)\big)^{1/2}$, where $V(r)$ is the gravitational potential. Therefore, when $r=0$, the observer clock rate is also zero; $R(r)=0$, which matches with the expected observation of the clock rate at the horizon: time freezes. As long as the gravitational field is uniform and weak, we can approximately comparing $1/\gamma=\sqrt{1-v^2/c^2}$ of special relativity to $R(r)\sim1-\big(V(r)/c^2\big)$ in general relativity, where $g_{tt}=-R^2(r)$. By a uniform field we mean it has one direction and constant value like the electric field between two infinitely long parallel plates. And by weak we mean $V(r)/c^2<<1$.  This means locally we can explain the redshift assuming  special relativity and the equivalence principle. But when the full theory of general relativity is assumed, things change. In special relativity you should detect redshift while the source is moving with constant speed from spatial point to another. This is Doppler shift and it's constant because it's characterized by the speed only. However in the presence of gravity and without source movement, we still can detect a gravitational redshift when the time-varying gravitational potential itself changes at the same point through time. The redshift is characterized by the change in the gravitational potential. So if we want to study the change in the gravitational potential, we need to consider some differential equation. If you want to solve this differential equation from SR viewpoint, you don't get correct answer. Rather you get Abraham theory or Nordstrom theory. And if you do the differential equation from GR viewpoint, i.e., general covariance differential equation, you get the answer that matches with experiments. But that answer is not compatible with SR as SR is based on inertial frames only. More details on this issue is reported here \cite{GRdisbute}.

In order to correlate how special relativity and general relativity see redshift, we should ask how acceleration/gravitation affect the clock rate. According to Don Koks \cite{Koks:1998}, we need to take into consideration what he calls is \emph{clock postulate}. The clock postulate generalizes the special relativistic comparison between the different clocks rates according to the Lorentz factor $\gamma$. This generalization is based on that even if the moving clock accelerates, the ratio of the rate of stationary clock compared to its rate is still scaled or compared by $\gamma$ as in relativistic Doppler effect.  Consequently, ratio depends only on $v$ not the derivatives of $v$.  Therefore, an accelerating clock counts out its time in a way that fore every one moment, its rate will slow by a factor $\gamma$ which is determined by instantaneous velocity at that moment; its acceleration has no effect at all. It is important to emphasize on that the clock postulate does not say that the counting out rate of a moving clock is unaffected by its acceleration. It says that the quantitative measurement of the relative rate does not depend on acceleration. Equivalently, the rate of such accelerated clock is indistinguishable from that of a clock in a \emph{momentarily comoving inertial frame}. For that, we can imagine an observer, in normal frame, is holding an inertial clock that for a brief moment slows to a stop alongside another accelerated observer with a clock, so that their relative velocity is momentarily zero. This is crucial as we shall see when we consider The Arnowitt-Deser-Misner (ADM) formalism.  At that moment they are ticking at the same rate. If that all is still confusing, you may consider the example of riding a bicycle on an icy morning that is mentioned in Ref. \cite{Koks:1998}. Contemplation about the clock postulate is not that obvious as the postulate can not be proved. Rather, simply it describes how we observe the physical world. It might be thought that the clock postulate mandates abandoning the Equivalence Principle as the Equivalence Principle equates between gravitational fields and acceleration. However, this is not true for the reasons explained in Ref. \cite{Koks:1998} by the example of a rocket empty of fuel and the two astronauts.

\section{One preferred frame}

\begin{figure}
    \centering
    \includegraphics[width=80mm]{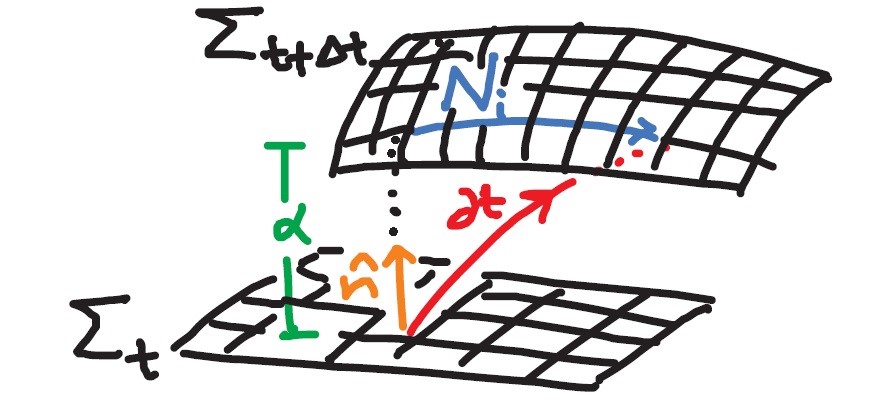}
    \caption{ADM slices of spacetime}
    \label{ADM}
\end{figure}

The absolute space and time means by definition a preferred frame, in which physics laws appear to be recognizably different from they appear in other frames. This Newtonian concept is contradicted with the assumed principle of relativity in inertial frames and the principle of covariance in general theory of relativity. However, all inertial frames are preferred over non-inertial frames in special relativity  as they observe only cause-and-effect relations between events in closed intervals, a property that general relativity lacks as it can observe a cause-and-effect relations between internal and external events. In this paper, we use the Rindler observer and the ADM formalism to restore the compatibility between the inertial privileged frames and relative space of general relativity, at least slice by slice.  For more details on this issue, see \cite{Wheeler,GRdisbute}. We consider spacetime foliation in ADM formalism \cite{Arnowitt:1962hi}, at every given moment according to canonical observers, each with its speed at that moment. These observers are related to each other by ``synchronicity'' \cite{LachiezeRey:2001gn}.
This approach comes with \textit{a global reference frame} that stays Minkowskian along the worldline of such observer. Despite that the absolute synchronicity is not attainable either in special relativity or in general relativity, a local synchronicity is extended beyond local neighborhoods as in Ref. \cite{LachiezeRey:2001gn}. This extension shows how the observed redshift can be useful in constructing a black hole universal clock with the help of relative gravitational redshift, as how we show in this letter, and the idea that the black hole is fully entangled with the space around it from Hawking radiation. The entanglement ``monogamy'' is saved by the cool horizon approach \cite{Maldacena:2013xja}. Like any clock, time will be measured in terms of lengths. This clock will form a  triangle that gives all measurements this clock can yield.  The Rindler observer \cite{Rindler:1966zz} would analyze the ADM formalism as if the shift function equal zero $N_i=0$. Therefore, the normal vector $\hat{n}$ to the spatial slice is proportional to the time basis $\partial_t$ of Rindler frame field with proportionality factor that is equal to the lapse function $\alpha$. In this case the coordinate observer is moving with velocity $(g_{tt})^{-1/2}e_t$ while the normal observer moves with velocity $\hat{n}$. The factor similar to $\gamma$ in ADM formalism is in form of $(1-N_iN^i/\alpha^2)^{-1/2}$. See Fig \ref{ADM}. Rindler observer, the previous factor is equal to 1. However this comparison between observers is NOT a physical relativist Lorentz boost, and the frame of reference can be dragged with any speed the observer chooses. It turns out that the normal observer and the Rindler  observer are \emph{the same} or \emph{One Observer}. They both see each other as if they are both belong to the same inertial Minkowski frame. This matches with the global reference frame described in Ref. \cite{LachiezeRey:2001gn}. The one observer idea may match with Wheeler idea of One-electron universe that he described to Feynman and on which Feynman depend to assume that "positrons could simply be represented as electrons going from the future to the past in a back section of their world lines" as Feynman stated in his  Nobel speech  \cite{Wheeler-Feynman}.

\section{The clock measurements}

 We assume an existence of Schwarzschild black hole with an event horizon. We investigate the gravitational red-shift which is a property of general covariance.  The Rindler's Observer measures the gravitational red-shift  which is a property of general covariance and its relative values between any two points or slices of spacetime. Therefore, we consider  two points ($A$ and $R$) in the gravitational field of black hole as shown in the following Fig. (\ref{blackholetriangle}). Notice here these two points form a triangle that follow a geodesics geometry of the considered black hole when connecting the two points with black hole center. If $R$ and $A$ are far enough from $K$, the triangle become approximately Euclidean triangle.
\begin{figure}[h!]
    \centering
    \includegraphics[width=80mm]{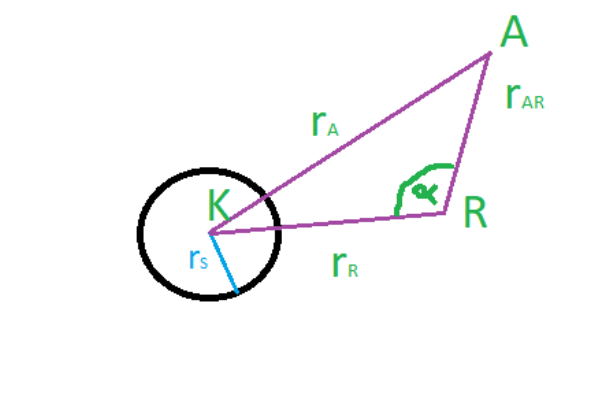}
    \caption{Black hole universal clock }
    \label{blackholetriangle}
\end{figure}
Between these two points $A$ and $R$, there are two possible clock measurements as follows:

\begin{enumerate}
    \item Relative gravitational red-shift which is represented by the ratio at two different points
    \begin{equation}
         \frac{z_A}{z_R}=\frac{(1-\frac{r_s}{r_A})^{-1/2}-1}
          {(1-\frac{r_s}{r_R})^{-1/2}-1}
          \label{Relative}
    \end{equation}
\item The difference in gravitational red-shift at two different points. 
    \begin{equation}
        \Delta z= z_A-z_R= (1-\frac{r_s}{r_A})^{-1/2}-(1-\frac{r_s}{r_R})^{-1/2}\label{Difference}
    \end{equation}
 \end{enumerate} 

\section{Weak Gravitational Approximation}
\subsection{Relative gravitational red-shift}
We consider the weak gravitational approximations, $r_s<<r_K$ and $r_s<<r_R$. The gravitational red-shift for both $A$ and $R$ can be approximated as follows
\begin{eqnarray}
     z_A&=&(1-\frac{r_s}{r_A})^{-1/2}-1\approx \frac{r_s}{2r_A} \\
z_R&=&(1-\frac{r_s}{r_R})^{-1/2}-1\approx \frac{r_s}{2r_R}.\label{shiftsAR}
\end{eqnarray}
We compute the relative gravitational red-shift using Eq(\ref{Relative}). We express it in terms of all lengths measured at $R$ including the distance between $A$ and $R$ ($r_{AR}$).

\begin{equation}
    \frac{z_R}{z_A}=\frac{1}{\sqrt{1-\frac{r_{AR}^2}{r_A^2}+2 \frac{r_R r_{AR}}{r_A^2} \cos{\alpha}}}=\delta
    \label{resolutionrelativity1}
\end{equation}
Notice the value of $\alpha$ can be $0 \leq \alpha \leq \pi/2$. This equation represents the relative gravitational-red-shift between two points $A$ and $R$ in a weak gravitational field. For the case $\alpha=\pi/2$. The relative gravitational red-shift will be given by

\begin{equation}
    \frac{z_R}{z_A}=\frac{1}{\sqrt{1-(\frac{r_{AR}}{r_A})^2}}=\delta \label{gammafactor}
\end{equation}
On the other side, the measurement in local inertial frames are determined in terms of relative time dilation as follows

\begin{equation}
    \frac{t_R}{t_A}=\frac{1}{\sqrt{1-\frac{v^2}{c^2}}}
\end{equation}
where $v$ is the relative speed between the two points $A$ and $R$ in the local inertial frames and $c$ is the speed of light. 

We notice the relative gravitational red-shift or relative gravitational time dilation  matches with the definition of time dilation in special relativity if the ratio $r_{AR}^2/{r_A}^2$ can be replaced by ratio $v^2/c^2$. The match is legitimate and mathematically consistent since the relative gravitational red-shift introduces a local measurement which is the time dilation in local inertial frames. Therefore, for mathematical consistency of general relativity, its local measurement should be equivalent to measurement special relativity that hold only in local inertial frames. This means  when $\alpha=\pi/2$, the gamma factor of special relativity emerges as a ratio between the gravitational red-shift at $A$ and  $R$. The clock measurements depends only on \textbf{``one variable''}; the distance from the gravitational source, which is the reason for velocity ratios turned to be lengths ratios in this delta factor in Eq. (\ref{gammafactor}). The ratio $r_{AR}^2/{r_A}^2$ can be considered as a geometric or gravitational interpretation of the ratio $v^2/c^2$. This comparison can be written as

\begin{equation}
    \frac{r_{AR}}{r_A}=\frac{r_{AR}/t}{r_A/t}=\frac{v}{c} \label{wholeandart}
\end{equation}
This would support the approach of time varying speed of light as a solution of cosmological puzzles that was suggested in  \cite{Albrecht:1998ir}. It may support also the experimental findings of changing physical constants such as fine structure constant in gravitational field as shown recently in \cite{Wilczynska:2020rxx}. In our case, the ratio $v/c$ varies depending on the distance from the gravitational source. 

We note that time can be inserted easily in the previous equation  as a ``redundant variable'' which  suggest a possible timeless state which is consistent mathematically. The timeless state  has been proposed in many contexts such as shape dynamics which introduce a gravitational origin of arrow of time 
\cite{Barbour:2011dn,Barbour:2013jya}. The timeless also emerged in Thermal time hypothesis
which assume that time only flow in thermodynamics or statistical patterns \cite{Connes:1994hv}.  It has been mathematically intuited  as well that timeless universe is possible  \cite{timelessmath}. In other words,  space may be a frozen time \cite{ibnarabi}.

To realize the effect of other  values of angle $\alpha$ in weak gravitational field, we consider an  approximation which is  $r_{AR}<<r_A$, $r_{AR}<<r_R$. In that case, the delta factor in Eq. (\ref{resolutionrelativity1}) is approximated as following

\begin{eqnarray}
      \delta \approx 1-\frac{r_R r_{AR}}{r_A^2}\cos{\alpha} \label{Kepler}
\end{eqnarray}
It is found that this equation matches with the derivative of Kepler equation. 

\begin{equation}
    \frac{dm}{dE}= 1-e \cos{E}
\end{equation}
where $m$ is the mean anomaly, $E$ is the eccentric anomaly, and $e$ is the eccentricity. In our approximation, the eccentricity $e$ is approximately equal to $r_R r_{AR}/r_A^2$, and $E$ refers to the angle $\alpha$. This gives a geometric interpretation of Kepler equation from the relative gravitational red-shift.

\subsection{Difference in Gravitational red-shift}

In this section, we compute the clock measurement as difference in gravitational red-shift. For weak gravitational approximation, we get

\begin{equation}
    \Delta z= z_R-z_A= \frac{r_s}{2r_R} -\frac{r_s}{2r_A} \label{differnce}
\end{equation}
Let  us make an approximation as following $r_A=r_R+x$, where $x<<r_A$ and $x<<r_R$. In that case, Eq. (\ref{differnce}) will be rewritten as follows. We use the value of Schwarzschild radius  $r_s=2 GM/c^2$

\begin{equation}
    \Delta z=G M \frac{x}{c^2 r_R^2} \label{mass1}
\end{equation}
where $G$, is the gravitational constant, $M$, is the black hole mass and $c$ is the speed of light. From Eq.(\ref{wholeandart}), $c$ can be set to equal to $r_A$ if we take $t$ to be unity since we agree that t is a redundant factor through matching local gravity measurement with local inertial frames.  We find that Eq. (\ref{mass1}) can be arranged to take the following form

\begin{equation}
    \Delta z~M=\Delta M=GM^2 \frac{x}{r_A^2 r_R^2} \label{emc3}
\end{equation}
where $\Delta M= \Delta z~M$. $\Delta M$ represents a possible mass between any two different points in the gravitational field. This would give a geometric representation for mass concept in terms of difference between different points to the black hole. We want to understand the physical meaning of the factor $G M^2$ in r.h.s of  Eq. (\ref{emc3}). When we look at  Bekenstein-Hawking entropy equation \cite{Bekenstein:1973ur,Hawking:1974sw}.

\begin{equation}
    S_{BH}= \frac{c^3 A}{4 G \hbar}=\frac{4 \pi}{c \hbar} G M^2
\end{equation}
where $A= 16 \pi (G M/c^2)^2$  stands for surface area of a black hole. We found that the factor $GM^2$ in r.h.s of  Eq. (\ref{emc3}) between any two different points can be expressed in terms of black hole entropy as follows

\begin{equation}
    \Delta z~M= \Delta M=\frac{\hbar}{4 \pi} \frac{x}{r_R^2 r_A} S_{BH} \label{mass}
\end{equation}
We assumed that time is a unity. let us consider this unit as the Planck time. This means that the Planck constant in previous equation can be replaced through the following process

\begin{equation}
    t_p=\sqrt{\frac{\hbar G}{c^5}}=1
\end{equation}
Since the Planck time is our unity, then c can set to be $r_A$. Therefore, the Planck constant in this geometric picture will be given by

\begin{equation}
    \hbar G= r_A^5
\end{equation}
This equation gives a geometric or gravitational interpretation of Planck constant. The relative mass between any two different points is therefore given by

\begin{equation}
    \Delta z M= \Delta M= \frac{1}{4 \pi G}\frac{x~ r_A^4}{r_R^2} S_{BH}=\frac{1}{16 \pi G}\frac{x ~~r_A^2}{r_R^2}~~ A \label{InformationMatter}
\end{equation}
The previous equation gives purely a geometric expression for the relative mass in terms of the gravitational source area of its full entropy. It is experimentally proved that the difference in gravitational potential has an effect on the apparent weight  of the 14.4-keV ray of Iron (Fe)  \cite{Pound:1960zz,Pound:1964zz}. This may be an experimental support for the derived relation that connect the difference in gravitational red-shift and emergence of mass in this section.



\section{Gravity and Uncertainty}

In previous sections, we have shown that the concept of velocity is replaced with the relative distance between any any two different points in the space time in the black hole universal clock. This would generate a timeless state  in a mathematically consistent way. In that state, the gravitational measurements happens in terms of only one variable which is the distance from the gravitational source. Time variable  appear to be a redundant variable. Since time, and therefore velocity dissolves, therefore there is no meaning to define uncertainty in this timeless state. We propose that the distance from the gravitational source may form the \emph{hidden variable} of every observation process in quantum mechanics. This may complete the connection between quantum mechanics and gravity in one unified theory, which is the timeless state. This may complete the picture that was introduced in EPR \cite{Einstein:1935rr}.
This implies that the uncertainty amount would decrease as the measurement happens closer to the gravitational source. The uncertainty \emph{emerges} due to the difference in information between point $A$ and point $R$ without knowing the distance to the source. This difference is encoded in Eq. \ref{InformationMatter}. The difference in information (uncertainty) would be represented by the difference in gravitational red-shift as follows:

\begin{equation}
    \frac{4 \pi \Delta M r_R^2 r_A}{x S_{BH}}= \hbar
\end{equation}
Notice that the difference in information between Point $A$ and point $R$ depends on the distances $r_A$ and $r_R$. If we do not know these values, this difference will be hidden in our local measurements, and therefore uncertainty emerges. Notice that the variables on the left hand side are greater than or equal to the Planck constant. This relation represent the hidden variables which is reason for emergence of  uncertainty principle inequality in local measurements. 

\section{Strong Gravity Case}

In strong gravity case, we can use Eqs. {\ref{Relative} and \ref{Difference}} without any approximation. These relations can be computed for any two points, and it gives a wide spectrum of measurements of relative gravitational red-shifts and masses in strong gravity field. In strong gravity field, the triangle will not be perfectly Euclidean but can be computed for every kind of measurement by knowing the length of this triangle.

\section{Why Clock Postulate/Timeless State of Gravity may be important?}

Time in principle is the change happening for any physical system. To approach a unified picture of physics, we can investigate  states in which no change happening, a frozen moment of the physical system. In that state, everything will be reduced to geometry. In this geometric picture, we can find a geometric picture of mass, speed of light, etc. Timeless state in that sense could introduce a unified picture of different concepts in physics. We hope to evolve this study in the future.    

\section{Conclusion}

We investigate how Rindler observer in ADM formalism make measurements in the slices of space-time. We found that the Rindler observer in ADM formalism  restores the compatibility between the inertial privileged frames and relative space of general relativity, at least slice by slice. We determined the relative or local measurements that Rindler observer could make in each slice of the space-time. We got a timeless state of the universe in which we found a geometric interpretations of speed of light and mass. The timeless state may form a correlation between relative gravitational red-shift and internal symmetries that are independent of time. We hope that this equivalence  may open the door for a gravitational technology. It is worth mentioning that timeless state of gravity can shed  a light on the nature of problem of time in Wheeler-De-Witt equation. We hope to report on these in the future.

\vspace{10 mm}

{\textbf Acknowledgment}
This work is supported by
the quantum gravity research grant, Los Angeles, California.

%


\end{document}